\documentclass[a4paper,11pt]{article}
\pdfoutput=1 

\usepackage{jheppub} 

\usepackage[T1]{fontenc} 

\usepackage{amsfonts,amsmath,amssymb}
\usepackage{enumerate}
\usepackage{hyperref}
\usepackage{graphicx}
\usepackage{caption}
\usepackage{subcaption}
\usepackage{ytableau}

\usepackage{enumitem}   

\usepackage{IEEEtrantools}
    
\newcommand{\be}{\begin{equation}}
\newcommand{\ee}{\end{equation}}

\newcommand{\bea}{\begin{eqnarray}}
\newcommand{\eea}{\end{eqnarray}}

\newcommand{\nn}{\nonumber}

\makeatletter
\newcommand{\citenosort}[1]{\begingroup\def\@addto@cite@list{\@cite@dump@now}\cite{#1}\endgroup}
\makeatother

\title{\boldmath  On the Superconformal Index of Chern-Simons theories and their KK Spectrometry}

\author[a]{Hyojoong Kim}
\author[a, b]{and Nakwoo Kim}

\affiliation[a]{Department of Physics 
and Research Institute of Basic Science, \\ Kyung Hee University, 
Seoul 02447, Republic of Korea}
\affiliation[b]{School of Physics, Korea Institute for Advanced Study, Seoul 02445, Republic of Korea}

\emailAdd{h.kim@khu.ac.kr}
\emailAdd{nkim@khu.ac.kr}

\abstract{We study the large-$N$ limit of superconformal index for two strongly interacting Chern-Simons theories in three dimensions with $\mathcal{N}=2$ supersymmetry, and compare the result against the AdS/CFT dual, namely the data of full Kaluza-Klein reduction spectra obtained using the exceptional field theory technique. The two theories of interest are mABJM and GJV theories, which are obtained as IR fixed point of RG whose UV description is ABJM and maximally supersymmetric Yang-Mills theory respectively. We confirm agreement and the duality persists. For the case of mABJM it turns out that we need a refinement of the gravity index which reflects the fact that the UV description on the field theory side has enhanced global symmetry.}

\begin{document} 
\maketitle
\flushbottom
\section{Introduction}
It has been known for a long time that explicit calculation of the Kaluza-Klein mass spectrum is a very difficult task in general backgrounds. The computations have been performed successfully only for a restricted class of background such as homogeneous Freund-Rubin type solutions.
However recently, starting with the seminal work of \cite{Malek:2019eaz, Malek:2020yue}, a new and powerful method to compute the Kaluza-Klein spectrum has been developed and applied to a number of examples. It turns out that this method can be applied to any vacuum in maximally gauged supergravity obtained through the consistent truncations of $D=10$ or $D=11$ supergravity \cite{Malek:2019eaz, Malek:2020yue, Malek:2020mlk, Varela:2020wty, Guarino:2020flh, Cesaro:2020soq, Bobev:2020lsk, Cesaro:2021haf}.\footnote{See \cite{Eloy:2020uix} for AdS$_3$ vacua in half-maximal gauged supergravity.}

The Kaluza-Klein spectrum plays an important role in the stability analysis.
The recent computation of the Kaluza-Klein mass spectrum of non-supersymmetric SO(3)$\times$ SO(3)-invariant AdS$_4$ vacuum in eleven-dimensional supergravity shows that there exist higher Kaluza-Klein modes whose mass-squared are below the Breitenlohner-Freedman bound. The result implies that this non-supersymmetric AdS$_4$ vacuum is perturbatively unstable \cite{Malek:2020mlk}. Similar analysis has been done for non-supersymmetric AdS$_4$ vacua of massive IIA theory \cite{Guarino:2020flh}.

Furthermore, in the context of AdS/CFT correspondence, the mass of the Kaluza-Klein modes can be mapped to the conformal dimension of the gauge invariant operators in dual field theory.
The analysis of the Kaluza-Klein spectrum of AdS$_5$ Pilch-Warner solution of type IIB supergravity enables us calculate the index over gravitons. It is shown that this graviton index exactly agrees with the superconformal index of four-dimensional $\mathcal{N}=1$ Leigh-Strassler superconformal field theory \cite{Bobev:2020lsk}.

In this work, we focus on two specific AdS solutions: ${\mathcal{N}}=2$ SU(3) $\times$ U(1)-invariant AdS$_4$ solutions in $D=11$ \cite{Corrado:2001nv} and massive IIA supergravity \cite{Guarino:2015jca}. The former can be obtained by uplifting Warner's AdS$_4$ solution \cite{Warner:1983vz} in ${\mathcal{N}}=8$ SO(8) gauged supergravity to eleven-dimensions.
Its dual is $D=3$ Chern-Simons-matter theory called mABJM theory, which can be obtained as an IR fixed point of the RG-flow triggered by a superpotential mass deformation of UV ABJM theory \cite{Benna:2008zy, Klebanov:2008vq}. 
 The latter solution can be obtained by uplifting the SU(3) $\times$ U(1)-invariant AdS$_4$ solutions of $D=4$, $\mathcal{N}=8$ dyonic ISO(7) gauged supergravity on $S^6$ \cite{Guarino:2015qaa, Guarino:2015vca}. Its dual field theory is called GJV theory, which can be obtained as an IR fixed point of the RG-flow triggered by a Chern-Simons deformation of $D=3$ UV SYM theory on D2-branes \cite{Guarino:2015jca}. 
  For these two sets of theories, the AdS/CFT correspondence has passed various non-trivial tests, for example, $S^3$-free energy, black hole entropy, etc.\footnote{See, for example, \cite{Jafferis:2011zi, Bobev:2018wbt, Kim:2019ewv, Bobev:2018uxk, Amariti:2021cpk, Guarino:2015jca, Hosseini:2017fjo, Benini:2017oxt}.}
  
  In this paper, utilizing the recent works on the Kaluza-Klein spectrum, we calculate the graviton index on 
${\mathcal{N}}=2$ SU(3) $\times$ U(1)-invariant AdS$_4$ solutions in $D=11$ and massive IIA supergravity.
On the field theory side, we compute the superconformal index, which was first discussed in $D=4$ superconformal field theories in \cite{Kinney:2005ej, Romelsberger:2005eg, Romelsberger:2007ec}.
For $D=3$,
the ${\mathcal N}=2$ superconformal index \cite{Bhattacharya:2008bja, Kim:2009wb, Imamura:2011su, Imamura:2011uj} is defined as
\be
I=\textrm{Tr} \left[\left(-1\right)^{F}\, e^{-\beta\{Q,S\}}\,x^{\Delta+j_3}\,\sum_i {z_i}^{F_i}\right].
\ee
Here $F$ is the fermion number and  $\Delta,\, R,\, j_3$ denote energy, R-charge, and the third component of the angular momentum, respectively. $z_i$ are fugacities for global symmetries and $F_i$ are the charges. Lastly, $Q$ and $S$ are certain supercharges in the ${\mathcal{N}}=2$ superconformal algebra, which satisfy
\be\label{BPS}
\{Q,S\}=\Delta-R-j_3 \geq 0.
\ee
Only the states saturating the above bound contribute to the index.
We show that the gravity calculations exactly agree with the field theory results. Hence, we provide one more precision test of AdS/CFT correspondence for mABJM and GJV theories, respectively.

This paper is organized as follows. In section \ref{mABJM}, we calculate the gravity and field theory index for mABJM theory. We show that they perfectly agree in the neutral sector and in the sector with one unit of magnetic flux. In section \ref{GJV}, we perform similar analysis for GJV theory. We conclude in section \ref{discussion}. In the appendix, we collect the various formulas that we use in the main part of this paper.
\section{mABJM theory}\label{mABJM}
The mABJM theory \cite{Benna:2008zy} is three-dimensional Chern-Simons-matter theory obtained by adding a superpotential mass term to one of the chiral multiplets of ABJM theory \cite{Aharony:2008ug}. This mass deformation triggers an RG-flow from UV ABJM theory to ${\mathcal{N}}=2$ superconformal field theory in the IR, which is called mABJM theory. Its gravity dual is given by an ${\mathcal{N}}=2$ AdS$_4$ solution of $D=11$ supergravity \cite{Corrado:2001nv}, which can be obtained by uplifting the SU(3) $\times$ U(1) invariant AdS$_4$ vacuum of ${\mathcal{N}}=8$ SO(8) gauged supergravity \cite{Warner:1983vz}. The correspondence of these two theories has passed various non-trivial tests, i.e., for example, matching the gauge theory operators with the supergravity multiplets \cite{Klebanov:2008vq}, an agreement of $S^3$-free energy \cite{Jafferis:2011zi, Bobev:2018wbt, Kim:2019ewv}, the topologically twisted index accounting for the entropy of black holes in AdS$_4$ \cite{Bobev:2018uxk} and, very recently, the holographic computation of $\tau_{RR}$ \cite{Amariti:2021cpk}.
In this section, we add to the list another non-trivial test of the correspondence, by checking the superconformal index. We compute the graviton index and the field theory index separately and show that two computations perfectly agree.
\subsection{Gravity index}
The $D=4,\, {\mathcal{N}}=2$ SU(3) $\times$ U(1) invariant AdS$_4$ solution and its uplift to $D=11$ supergravity were constructed a long time ago in \cite{Warner:1983vz} and in \cite{Corrado:2001nv}, respectively.
However, its complete Kaluza-Klein spectrum was not known. Only a small subset of the spectrum, for example,
the spectrum of the short multiplets \cite{Klebanov:2008vq}
and the spin-2 spectrum \cite{Klebanov:2009kp} was known. It was only recently that the complete spectrum was successfully calculated based on the Exceptional Field Theory techniques in \cite{Malek:2019eaz, Malek:2020yue}.

In this paper, we focus on the spectrum of the short multiplets\footnote{We follow the conventions used in \cite{Klebanov:2008vq}.}, even though the full spectrum is available, because we will compute the gravity index and compare it to the superconformal index of dual field theory. The spectrum of the short multiplets is summarized in table \ref{table:mABJM}, which contains all the information needed for the index: the names of the short multiplets, their SU(3) $\times$ U(1)$_r$ representations and the energy $E_0$. Here $[p,q]$ represents the SU(3) Dynkin label, $r$ is the R-charge and $n$ is the Kaluza-Klein level. See table 6 in \cite{Klebanov:2008vq} and equations (5.66) and (5.67) in \cite{Malek:2020yue}.

\begin{table}[h!]
\begin{center}
\begin{tabular}{l|ll||l}
 & \hfil $[p,q]_r$ & $\hfil E_0$ & \hfil index\\
\hline
 SGRAV & $[0,0]_{\pm n}$ & $E_0=n+2$ & $-x^{4+n}$ \\
 SGINO & $[n+1,0]_{(n+1)/3} \oplus [0,n+1]_{-(n+1)/3}$ & $E_0=\frac{11}{6}+\frac{n}{3}$ 
& $\phantom{+}x^{\frac{10+n}{3}}$\\
 SVEC & $[n+1,1]_{n/3} \oplus [1,n+1]_{-n/3}$ & $E_0=\frac{n+3}{3}$ & $-x^{\frac{6+n}{3}}$\\ 
 HYP & $[n+2,0]_{(n+2)/3} \oplus [0,n+2]_{-(n+2)/3}$ & $E_0=\frac{n+2}{3}$ & 
$\phantom{+}x^{\frac{2+n}{3}}$\\
\end{tabular}
\caption{The spectrum of the short multiplets at the Kaluza-Klein level $n$ and the contributions to the superconformal index for ${\mathcal{N}}=2$ SU(3) $\times$ U(1) invariant AdS$_4$ solution in $D=11$ supergravity.}
\label{table:mABJM}
\end{center}
\end{table}
Here SGRAV, SGINO, SVEC, HYP represent ${\mathcal{N}}=2$ short graviton, short gravitino, short vector and hyper multiplet, respectively. For $n=0$, SGRAV and SVEC reduce to a massless graviton multiplet MGRAV and a massless vector multiplet MVEC, respectively. We compute the contributions to the index for given short multiplets and write them down in the last column of table \ref{table:mABJM}, which will be multiplied by the $SU(3)$ characters. Let us explain in more detail. The fields in the supermultiplets are labeled by $(s, E, y)$, where spin $s$, energy $E$ and hypercharge $y$.\footnote{See table 8 $\sim$ 16 in \cite{Klebanov:2008vq}. Tables 9, 11, 14, 16 are relevant to our calculations in this section. See also the section 4.2 of \cite{Cordova:2016emh}.} For SGRAV multiplet, the fields with $(\frac{3}{2}, E_0+\frac{1}{2}, y_0+1)$ saturate the bound \eqref{BPS} and contribute $-x^{4+n}$ to the index. Similarly, the fields labeled by $(1, E_0+\frac{1}{2}, y_0+1)$ in SGINO, $(\frac{1}{2}, E_0+\frac{1}{2}, y_0+1)$ in SVEC and $(0, E_0, y_0)$ in HYP contribute to the index by $x^{\frac{10+n}{3}},\, -x^{\frac{6+n}{3}}$ and $x^{\frac{2+n}{3}}$, respectively. On top of that, we should multiply the specific SU(3) characters for given representations of the supermultiplets. Only multiplets with $r \geq 0$ in the above table contribute to the index. Taking into account the multiples with spacetime derivatives, one should multiply an overall factor of $\frac{1}{1-x^2}$. Finally, we sum the contributions over all the Kaluza-Klein levels and obtain the single-graviton index as
\begin{align}
&I_{\textrm{mABJM}}^\textrm{sp}\left(x,y_1,y_2 \right)\nn\\
&=\dfrac{1}{1-x^2}\sum_{n=0}^{\infty} \left( - x^{4+n}\, \chi_{SU(3)}^{[0,0]}
+ x^{\frac{10+n}{3}} \, \chi_{SU(3)}^{[n+1,0]}
- x^{\frac{6+n}{3}} \, \chi_{SU(3)}^{[n+1,1]}
+ x^{\frac{2+n}{3}} \, \chi_{SU(3)}^{[n+2,0]}
\right)\label{mABJM-gravity0}\\
&\phantom{=}\,\,
 +\dfrac{1}{1-x^2}
 \left(
- x^{\frac{5}{3}}  \chi_{SU(3)}^{[0,1]}
 +x^{\frac{1}{3}}  \chi_{SU(3)}^{[1,0]}
 \right),\nn\\
&=
\dfrac{
\left(1-  {y_1}^{-1} x^{\frac{5}{3}} \right)
\left(1-y_1 {y_2}^{-1} x^{\frac{5}{3}}  \right)
\left(1-y_2 x^{\frac{5}{3}}  \right)
}{
\left(1- y_1  x^{\frac{1}{3}} \right)
\left(1-{y_1}^{-1} y_2  x^{\frac{1}{3}} \right)
\left(1-{y_2}^{-1} x^{\frac{1}{3}}  \right)
\left(1-x^2\right)^2}-\dfrac{1-x^2+x^4}{(1-x^2)^2}\label{mABJM-gravity} . 
\end{align}
Here $\chi_{SU(3)}^{[p,q]}$ is the SU(3) character with Dynkin labels $[p,q]$, whose explicit form is written in \eqref{chi-su(3)}. In the second line of equation \eqref{mABJM-gravity0}, we have added the contributions from SVEC and HYP with $n=-1$. They correspond to the contribution from the spin $\frac{1}{2}$ and $0$ ultra-short singleton supermultiplets \cite{Gunaydin:1985tc} sitting at the bottom of the Kaluza-Klein tower.
The singleton fields are known to be free and live on the boundary. Hence, they are decoupled from the spectrum. In the viewpoint of AdS/CFT correspondence, they correspond to the decoupled U(1) sector of dual field theory.
See, for example, \cite{Gunaydin:1998km, Witten:1998qj, Maldacena:2001ss}.

This index can be also obtained from the ABJM index, i.e. the single-graviton index on AdS$_4 \times$ S$^7$ \eqref{ABJM-index} by substituting 
\be
y_1\rightarrow x^{\frac{1}{3}}y_1, \quad y_2\rightarrow x^{-\frac{1}{3}}y_2, \quad y_3\rightarrow x^{\frac{1}{3}}y_2 / y_1.
\ee
It reflects the fact that ABJM theory and mABJM theory are related by an RG-flow.\footnote{For AdS$_5$/CFT$_4$ correspondence, similar analysis was carried out between ${\mathcal{N}}=4$ SYM theory in the UV and the Leigh-Strassler SCFT in the IR \cite{Bobev:2020lsk}.}
\subsection{Superconformal index}
In this section, we compute the superconformal index of mABJM theory, which can be obtained as an IR fixed point of an RG-flow starting from UV ABJM theory \cite{Benna:2008zy, Klebanov:2008vq}. More specifically, the flow is triggered by a mass term
\be
\Delta W \sim \textrm{Tr} \left(T^{(1)} A_1 \right)^2,
\ee
to the superpotential of $k=1, 2$ ABJM theory. Here, for concreteness, we choose one specific chiral multiplet $A_1$. $T^{(1)}$ is the monopole operator in $(\mathbf{N},\mathbf{\bar{N}})$ representation. This superpotential deformation preserves ${\mathcal{N}}=2$ supersymmetry, but breaks SU(4)$_R \times $ U(1)$_b$ symmetry of ${\mathcal{N}}=6$ ABJM theory to SU(3)$_F \times$ U(1)$_R$. 

Let us begin by summarizing various charges of the bosonic fields in ABJM and mABJM theories in the table below.\footnote{We follow the convention of \cite{Kim:2009wb} for charges of ABJM theory. For the charges of the fermionic fields and supercharges, we refer to this reference.} 

\begin{center}
\begin{tabular}{c|c c c c c c|c c c}
 &  & & ABJM & & & & &mABJM & \\
\hline
fields & $\phantom{-}h_1$ & $\phantom{-}h_2$ & $\phantom{-}h_3$ & $\phantom{-}h_4$ & $j_3$ & $\Delta$ & $\phantom{-}t_1$ & $\phantom{-}t_2$ & $\Delta$\\
\hline
$A_1$  & $\phantom{-}\frac{1}{2}$ & $\phantom{-}\frac{1}{2}$ & $-\frac{1}{2}$& $\phantom{-}\frac{1}{2}$ & 0 & $\frac{1}{2}$ & $\phantom{-}0$             &   $\phantom{-}0$                     & $1$           \\ 
$A_2$  & $-\frac{1}{2}$ & $-\frac{1}{2}$ & $-\frac{1}{2}$ & $\phantom{-}\frac{1}{2}$ & 0 & $\frac{1}{2}$ & $\phantom{-}0$             &   $\phantom{-}\frac{1}{2}$   & $\frac{1}{3}$ \\ 
$B_1$  & $\phantom{-}\frac{1}{2}$ & $-\frac{1}{2}$ & $-\frac{1}{2}$ & $-\frac{1}{2}$ & 0 & $\frac{1}{2}$ & $\phantom{-}\frac{1}{2}$   &   $-\frac{1}{2}$   & $\frac{1}{3}$ \\ 
$B_2$  & $-\frac{1}{2}$ & $\phantom{-}\frac{1}{2}$ & $-\frac{1}{2}$ & $-\frac{1}{2}$ & 0 & $\frac{1}{2}$ & $-\frac{1}{2}$  &   $\phantom{-} 0$   & $\frac{1}{3}$ \\ 
\end{tabular}
\end{center}

The superconformal symmetry of ABJM theory is Osp(6|4), whose bosonic subgroup is SO(6) $\times$ SO(3,2).
Here, $h_1, h_2, h_3$ and $j_3, \Delta $ are the charges of the Cartans in SO(6) $\times$ SO(3) $\times$ SO(2). $\Delta$ is the scaling dimension of the fields and equal to $\frac{1}{2}$.
$h_4$ is a charge of a baryon-like U(1)$_b$ symmetry.
In ABJM theory, it is known that the monopole operator plays a central role.
The monopole operators are charged under U(1)$_b$ symmetry and turns on magnetic fluxes through an $S^2$ surrounding the point where the monopole operator is inserted. Constructing the gauge invariant operators dressed with the monopole operators, one can show that U(1)$_R \times$ SU(2) $\times$ SU(2) $\times$ U(1)$_b$ symmetry, which is manifest with the above charge assignments, is enhanced to SU(4) $\times$ U(1)$_b$.

The superconformal index of ABJM theory was first computed in the neutral sector, where the monopole operator is not included. The field theoretic result matches the gravitational calculation on AdS$_4 \times S^7/\mathbb{Z}_k$ in the large $k$ 't Hooft limit \cite{Bhattacharya:2008bja}. After including the monopole operators, the index in the large $N$ limit shows a perfect agreement with the graviton index on AdS$_4 \times S^7/\mathbb{Z}_k$ with an arbitrary $k$ \cite{Kim:2009wb}.
Then, the index computation was generalized to other ${\mathcal{N}}=2$ quiver gauge theories with arbitrary R-charges \cite{Imamura:2011su,Imamura:2011uj,Cheon:2011th}.

Now let us turn to the mABJM theory. The symmetry of mABJM theory is SU(3) $\times$ Osp(2|4), whose bosonic subgroup is SU(3) $\times$ U(1)$_R$ $\times$ SO(3,2).
Here $t_1$ and $t_2$ are the charges of the Cartans in SU(3). The scaling dimensions of the fields at the IR fixed point are different than those of the free fields. 
The monopole operator also plays an important role in constructing the gauge invariant operators in mABJM theory \cite{Klebanov:2008vq}. Without it, only the SU(2) symmetry is seen manifestly. This symmetry is enhanced to SU(3) after introducing a monopole operator. However, note that, in contrast to ABJM theory, the baryon-like U(1)$_b$ on its own is not a symmetry in mABJM theory. Its presence in the symmetry algebra is only as a linear combination with another U(1) whose charge we denote by $h_1$, thereby leading to an element in the Cartan subalgebra of SU(3). It can be easily seen if one rewrites the Cartan charges in terms of the charges of ABJM theory as $t_1=\left(h_1-h_2 \right)/2,\, t_2=-\left(h_1-h_4 \right)/2$.

Now let us consider the superconformal index of mABJM theory. 
We follow the prescription of Romelsberger  \cite{Romelsberger:2007ec, Gadde:2010en} and calculate the index of the UV description with the R-charge assignments of the IR theory.
The index can be calculated from the path integral on $S^2 \times S^1$, which localizes at the saddle points labeled by magnetic fluxes on $S^2$ and the holonomy zero modes along $S^1$. 
We closely follow the method developed in \cite{Kim:2009wb, Imamura:2011su} and summarize various formulas needed in the computations in the appendix \ref{review-SCI}. 
 For the case at hand, the U(1)$_b$ is not a symmetry of mABJM theory. Therefore, we compute the index with the fugacities $z_1, q$ and $q_b$ for charges $t_1, -h_1/2$ and $h_4/2$ instead of the fugacities $z_1, z_2$ for $t_1, t_2$. Then, we identify $q$ and $q_b$ later.
 
In this paper we are interested in comparing the field theory index to the gravitational one. Hence, we focus on the large $N$ limit of superconformal index. At large $N$, it is known that the index is factorized into
\be
I=I^{(0)}I^{(+)}I^{(-)},
\ee
where $I^{(0)}$ is the index without the monopole operators and $I^{(+)}\left(I^{(-)}\right)$ represents the index with the positive(negative) magnetic fluxes \cite{Kim:2009wb}.

First, let us calculate the neutral part of the index. Without monopole operators, the theory does not enjoy the full SU(3) symmetry, but only exhibits SU(2). Hence, we consider the index with the fugacity $z_1$. Then, one can easily read off the matrix $M$ in \eqref{fprime} and obtain the neutral part of the index \eqref{neutral-index} as
\begin{align}\label{SCI-mABJM-neutral}
I^{(0)}
=
\prod_{n=1}^{\infty}\dfrac{(1-x^{2n})^2}{
\left(1-{z_1}^{-\frac{n}{2}} x^{\frac{2n}{3}}\right)
\left(1-{z_1}^{\frac{n}{2}}  x^{\frac{2n}{3}}\right)
\left(1-{z_1}^{-\frac{n}{2}} x^{\frac{4n}{3}}\right)
\left(1-{z_1}^{\frac{n}{2}} x^{\frac{4n}{3}}\right)}.
\end{align}
We also calculate the index with the fugacities $z_1, q$ and $q_b$. The neutral part of the index is independent of $q_b$, as expected. Then by setting $q$ to one, we obtain the above result.

We move on to the computation of the index of the sector with the monopole operators. As illustrated in \cite{Kim:2009wb, Imamura:2011su}, we will consider 
Dirac monopoles in U(1)$^N \times$ U(1)$^N \subset$ U(N) $\times$ U(N).
Here, we restrict ourselves to the diagonal magnetic monopole. It implies that its magnetic charges satisfy the relation $\sum_i m_i =\sum_i \tilde{m}_i$. 
 For simplicity, we compute the index in the sector with the unit magnetic flux as
\be
m=\left(1,0,\cdots,0 ; 1,0,\cdots,0\right).
\ee
Its positive part and the corresponding holonomy can be denoted as
\be
m^{(+)}=(1 ; 1), \quad a^{(+)}=(a; \tilde{a}).
\ee
Following the procedures summarized in the appendix \ref{review-SCI}, we calculate the zero-point contributions  and the classical contribution from the Chern-Simons action as
\be
\epsilon_0^{(+)}=0, \quad q_{0i}^{(+)}=0, \quad S_{CS}^{(+)}= i k (a-\tilde{a}).
\ee
Then, the index is given by
\be
I_{\ytableausetup{boxsize=0.3em}\ydiagram{1}\,\,\ydiagram{1}}^{(+)}=
\int \dfrac{da}{2\pi}\, \dfrac{d\tilde{a}}{2\pi}\,e^{-i k (a-\tilde{a})}\,\textrm{exp}\left[\sum_{n=1}^{\infty} \dfrac{1}{n}f^{+}\left(e^{i n(a-\tilde{a})}, {z_1}^n, q^n, {q_b}^n, x^n\right)\right],
\ee
where the letter index from the vector and the chiral multiplets \eqref{f-vec}, \eqref{f-chi} are
\begin{align}
f^{(+)}_{\textrm{vector}}&=2x^2,\\
f^{(+)}_{\textrm{chiral}}
&=
\sum_{A_i}\left( e^{i(a-\tilde{a})} {z_1}^{t_1} q^{-\frac{h_1}{2}} {q_b}^{\frac{h_4}{2}} x^{\Delta} 
-e^{-i(a-\tilde{a})} {z_1}^{-t_1} q^{\frac{h_1}{2}} {q_b}^{-\frac{h_4}{2}}x^{2-\Delta} \right)\nn\\
&+\sum_{B_i}\left( e^{i(\tilde{a}-a)} {z_1}^{t_1} q^{-\frac{h_1}{2}} {q_b}^{\frac{h_4}{2}} x^{\Delta} 
-e^{-i(\tilde{a}-a)} {z_1}^{-t_1} q^{\frac{h_1}{2}} {q_b}^{-\frac{h_4}{2}}x^{2-\Delta} \right).
\end{align}
Calculating the plethystic exponential and defining an integration variable $z\equiv e^{i \left(a-\tilde{a} \right)}$, we obtain
\begin{align}\label{SCI-mABJM-pre}
I_{\ytableausetup{boxsize=0.3em}\ydiagram{1}\,\,\ydiagram{1}}^{(+)} \left(x, z_1, q \right)
= &\oint
\dfrac{dz}{(2\pi i)z}\, z^{-k} \\
&\times\left[
\dfrac{
\left(1-{z_1}^{-\frac{1}{2}} q^{\frac{1}{4}}z x^{\frac{5}{3}} \right)
\left(1-{z_1}^{\frac{1}{2}} q^{-\frac{1}{4}}z x^{\frac{5}{3}}\right)
\left(1- q^{-\frac{1}{4}} z^{-1} x^{\frac{5}{3}}\right)
\left(1- q^{\frac{1}{4}}z^{-1} x \right)}{
\left(1-{z_1}^{\frac{1}{2}}  q^{-\frac{1}{4}}z^{-1} x^{\frac{1}{3}} \right)
\left(1-{z_1}^{-\frac{1}{2}} q^{\frac{1}{4}}z^{-1} x^{\frac{1}{3}} \right)
\left(1- q^{\frac{1}{4}} z x^{\frac{1}{3}} \right)
\left(1- q^{-\frac{1}{4}}z x\right)
\left(1-x^2\right)^2}\right].\nn
\end{align}
Here the fugacity ${q_b}$ is absorbed in $z$ so that $z$ plays a role of ${q_b}^{\frac{1}{4}}$ in the bracket in \eqref{SCI-mABJM-pre}. This function in the bracket can be obtained by substituting 
\be
y_1 \rightarrow {t_1}^{\frac{1}{2}} q^{-\frac{1}{2}} x^{\frac{1}{3}}, \quad
y_2 \rightarrow  {t_1}^{\frac{1}{2}}  x^{-\frac{1}{3}}, \quad
y_3 \rightarrow z^2 x^{\frac{1}{3}}
\ee
into the function $F(x, y_1, y_2, y_3)$ defined in the ABJM theory \cite{Kim:2009wb}.\footnote{See equation (C.3) in \cite{Kim:2009wb}.}

The quantity  $I_{\ytableausetup{boxsize=0.3em}\ydiagram{1}\,\,\ydiagram{1}}^{(+)}$, which we have just calculated in \eqref{SCI-mABJM-pre}, is the contribution to the index from the sector with unit magnetic flux. By summing up all the contributions from the various magnetic flux configurations, one can obtain the index $I^{(+)}$ as
\be
I^{(+)} \left(x, z_1, q, z  \right)=1+z^k\, I_{\ytableausetup{boxsize=0.3em}\ydiagram{1}\,\,\ydiagram{1}}^{(+)} \left(x, z_1, q \right)+ \cdots.
\ee
In other words, once we know the index $I^{(+)} \left(x, z_1, q,z  \right)$, then we can expand it in $z$ and obtain $I_{\ytableausetup{boxsize=0.3em}\ydiagram{1}\,\,\ydiagram{1}}^{(+)}$ from the coefficient of $z^k$. However, it is not possible to calculate $I^{(+)}$ by summing up all the monopole operator contributions in general. 
Hence, instead of comparing $I^{(+)}$ and $I^{(+)}_\textrm{mp}$, we compare the field theory calculation and the gravity result for each sector, for example, the sector with unit magnetic flux as we will consider in this paper.

Furthermore, since we consider the index with SU(3) symmetry, we identify $q$ and $q_b$, and denote them as $z_2$, as we mentioned earlier. It leads us to identify $z \equiv {z_2}^{\frac{1}{4}}$ and obtain
\be
I^{(+)} \left(x, z_1, z_2 \right)=1+{z_2}^{\frac{k}{4}}\, I_{\ytableausetup{boxsize=0.3em}\ydiagram{1}\,\,\ydiagram{1}}^{(+)} \left(x, z_1, z_2 \right)+ \cdots.
\ee
Once we identify $q$ and $q_b$, and obtain the index associated with SU(3) symmetry, we are not allowed to expand the index in terms of fugacity associated to the monopole operator any more.
\subsection{Refined gravity index}
In previous sections, we have computed the single-graviton index and the field theory index. More specifically, the field theory index has been calculated in the sector with zero and one monopole operator. To compare these gravity and field theory results, let us revisit the gravity index in this section.

In the class of theories having U(1)$_b$ symmetry,\footnote{It includes ABJM theory \cite{Bhattacharya:2008bja,Kim:2009wb} 
and examples studied in \cite{Cheon:2011th}.}
 one expands the gravity index in terms of the fugacity associated to U(1)$_b$ symmetry, which is realized as the isometry along the Hopf fibration in seven-dimensional internal manifolds.
 Then, one may reproduce the field theory results, i.e. the neutral sector index which is not charged under U(1)$_b$ and the index in the sector with monopole operators. However, mABJM theory does not have a U(1)$_b$ symmetry: it has only SU(3) flavor symmetry. As a result, the gravity index is written in terms of the SU(3) characters. Hence, it is not clear what should be the expansion parameter which allows us to distinguish the index in the neutral sector and in the sector with monopole operators, in the gravity result \eqref{mABJM-gravity}. Here, we point out that we need more information than given in the single-graviton index  \eqref{mABJM-gravity} in terms of SU(3) representations. 

Let us begin with the single-graviton index in AdS$_4 \times S^7$ \eqref{ABJM-gravity-su4}, which is written in terms of SU(4) characters. We rewrite this expression using U(3) characters $\chi_{U(3)}^{[p,q]}(x_1, x_2, x_3)$ and substitute 
\be
x_1 \rightarrow x^{-\frac{1}{6}} x_1, \quad
x_2 \rightarrow x^{-\frac{1}{6}} x_2, \quad
x_3 \rightarrow x^{-\frac{1}{6}} x_3, \quad
x_4 \rightarrow x^{\frac{1}{2}} x_4,
\ee
where $x_1 x_2 x_3 x_4=1$. Then, we have
\begin{IEEEeqnarray}{ll}
\IEEEeqnarraymulticol{2}{l}{
(1-x^2)\tilde{I}_{\textrm{mABJM}}^{\textrm{sp}}(x, x_1, x_2,x_3)}\label{refined}\\
=\sum_{n=0}^{\infty} 
\Bigg\{
&-x^{\frac{n}{2}+4}  \left( \textcolor{blue}{x^{\frac{n}{2}} \chi_{U(3)}^{[0,0]} x_4^n}
+\sum_{i=0}^{n-1} x^{-\frac{n-4i}{6}} \chi_{U(3)}^{[n-i,0]} x_4^i\right)\nn\\
&+x^{\frac{n}{2}+3} \left( \textcolor{blue}{x^{-\frac{n-2}{6}} \chi_{U(3)}^{[n+1,0]} x_4 }
+\sum_{i=0}^{n-1} x^{-\frac{n-4i}{6}+1} \chi_{U(3)}^{[n-i,0]} x_4^{i+2}
+\sum_{i=-1}^{n-1} x^{-\frac{n-2-4i}{6}} \chi_{U(3)}^{[n-i-1,1]} x_4^{i+1}
 \right) \nn\\
&-x^{\frac{n}{2}+2} \Bigg( \textcolor{blue}{x^{-\frac{n}{6}} \chi_{U(3)}^{[n+1,1]} x_4} \nn\\
 &\phantom{-x^{\frac{n}{2}+2} \Bigg(}\left.
 + \sum_{i=-1}^{n-1} x^{-\frac{n-2-4i}{6}+1} \chi_{U(3)}^{[n-i-1,1]} x_4^{i+3}
 +\sum_{i=1}^{n+2} x^{-\frac{n+2-4i}{6}-1}  \chi_{U(3)}^{[n+2-i,0]} x_4^{i-2}
 \right) \nn\\
 &+x^{\frac{n}{2}+1} \left( \textcolor{blue}{x^{-\frac{n+2}{6}} \chi_{U(3)}^{[n+2,0]} }
 +\sum_{i=1}^{n+2} x^{-\frac{n+2-4i}{6}} \chi_{U(3)}^{[n+2-i,0]} x_4^{i} \right)\Bigg\}\nn\\
\IEEEeqnarraymulticol{2}{l}{
 \phantom{=}-x^{\frac{3}{2}} \left(  \textcolor{blue}{x^{\frac{1}{6}} \chi_{U(3)}^{[0,1]} x_4 }
+x^{-\frac{1}{2}}x_4^{-1}\right)
+x^{\frac{1}{2}} \left( \textcolor{blue}{x^{-\frac{1}{6}} \chi_{U(3)}^{[1,0]}}+x^{\frac{1}{2}}x_4\right).}\nn
\end{IEEEeqnarray}
Here we use $\tilde{I}_{\textrm{mABJM}}^{\textrm{sp}}$ to denote a refined version of the single-graviton index with one more fugacity compared to the original single-graviton index calculated in \eqref{mABJM-gravity0}. 
When we impose $x_1 x_2 x_3=1$, or equivalently $x_4=1$, all the U(3) characters in the above expression reduce to SU(3) characters. Then, the terms written in blue in \eqref{refined} exactly match the index which is calculated directly from the KK-spectroscopy analysis in \eqref{mABJM-gravity0}. The remaining terms completely cancel with each other and do not contribute to the index in this case.

Now we are ready to reproduce the field theory results presented in the previous section. To do that, it is more appropriate to use the following expression
\begin{align}\label{refined-sum}
\tilde{I}_{\textrm{mABJM}}^{\textrm{sp}}=
\dfrac{(1-x_1^{-1}x^{\frac{5}{3}})(1-x_2^{-1}x^{\frac{5}{3}})(1-x_3^{-1}x^{\frac{5}{3}})
(1- x_1 x_2 x_3 x)}
{(1-x_1 x^{\frac{1}{3}})(1-x_2 x^{\frac{1}{3}})(1-x_3 x^{\frac{1}{3}})
(1-(x_1 x_2 x_3)^{-1}x)(1-x^2)^2}
-\dfrac{1-x^2+x^4}{(1-x^2)^2},
\end{align}
which can be obtained by evaluating the summations in \eqref{refined}.
We identify the fugacities of the gravity and field theory indices as
\be
x_1 \equiv {z_1}^{\frac{1}{2}}  q^{-\frac{1}{4}}z^ {-1}, \quad
x_2 \equiv {z_1}^{-\frac{1}{2}} q^{\frac{1}{4}}z^{-1}, \quad
x_3 \equiv q^{\frac{1}{4}} z,
\ee
by comparing the refined single-graviton index \eqref{refined-sum} and the field theory index in the sector with one monopole operator \eqref{SCI-mABJM-pre}. With this identification, we can rewrite the refined gravity index in terms of the field theory fugacities $z_1, q, z$. We already know from the field theory analysis that $z$ plays the role of $q_b^{\frac{1}{4}}$, which is the fugacity associated with the baryonic U(1)$_b$. Hence, it implies that we finally obtain the expression which we can expand in $z$, i.e. the fugacity associated to the monopole operators and reproduce the field theory results.

Before doing that, we mention that the refined index \eqref{refined-sum} reduces to the original one by imposing the constraint $x_1 x_2 x_3 =1$. More specifically, we can reproduce \eqref{mABJM-gravity} with
\be
x_1= y_1, \quad x_2=\dfrac{1}{y_2}, \quad x_3 =\dfrac{y_2}{y_1}.
\ee
The constraint $x_1 x_2 x_3 =1$ translates into the identification $z \equiv q^{\frac{1}{4}}$ on the field theory side. 

Now we check the agreement between the gravity and field theory calculations in each sector. First, we focus on the neutral sector index.
We can calculate the neutral sector index by expanding $\tilde{I}_{\textrm{mABJM}}^{\textrm{sp}}\left(x, z_1, q, z \right)$ in $z$ and read off the $z$-independent part. It can be done by evaluating the contour integral
\begin{align}
\oint
\dfrac{dz}{(2\pi i)z}  \tilde{I}_{\textrm{mABJM}}^{\textrm{sp}}\left(x, z_1, q, z \right),
\end{align}
which includes the poles at $z=0,\, z={z_1}^{-\frac{1}{2}} q^{\frac{1}{4}} x^{\frac{1}{3}},\,z={z_1}^{\frac{1}{2}} q^{-\frac{1}{4}} x^{\frac{1}{3}}$. As a result, we obtain the single particle index as
\begin{align}
\tilde{I}_{\textrm{mABJM}}^{\textrm{sp}}(x, z_1, q)=
&-2-\dfrac{2}{1-x^2}\nn\\
&+\dfrac{1}{1-{z_1}^{-\frac{1}{2}} q^{\frac{1}{2}}x^{\frac{2}{3}}}
+\dfrac{1}{1-{z_1}^{\frac{1}{2}} x^{\frac{2}{3}}}
+\dfrac{1}{1-{z_1}^{-\frac{1}{2}} x^{\frac{4}{3}}}
+\dfrac{1}{1-{z_1}^{\frac{1}{2}} q^{-\frac{1}{2}}x^{\frac{4}{3}}}.
\end{align}
After setting $q=1$, we calculate the multi-particle index as
\be
\tilde{I}_{\textrm{mABJM}}^{\textrm{mp}}(x, {z_1})=\textrm{exp}\left[\sum_{n=1}^{\infty}\dfrac{1}{n}\tilde{I}_{\textrm{mABJM}}^{\textrm{sp}}(x^n, {z_1}^n)\right],
\ee
and successfully reproduce the field theory index in the neutral sector \eqref{SCI-mABJM-neutral}.

Let us move on to the index in the sector with one monopole operator. In this sector, we also expand 
$\tilde{I}_{\textrm{mABJM}}^{\textrm{sp}} \left(x, z_1, q,z \right)$ in $z$ and compute the coefficient of $z^k$ as
\begin{align}
\tilde{I}_k^\textrm{sp} \left(x, z_1, q \right)
= \oint
\dfrac{dz}{(2\pi i)z}\, z^{-k}\,\tilde{I}_{\textrm{mABJM}}^{\textrm{sp}}\left(x, z_1, q, z \right).
\end{align}
The field theory index \eqref{SCI-mABJM-pre} can be neatly rewritten as
\be\label{index-one-mono}
I_{\ytableausetup{boxsize=0.3em}\ydiagram{1}\,\,\ydiagram{1}}^{(+)} \left(x, z_1, q \right)
= \oint
\dfrac{dz}{(2\pi i)z}\, z^{-k} \left ( \tilde{I}_{\textrm{mABJM}}^{\textrm{sp}}\left(x, z_1, q, z \right)+\dfrac{1-x^2+x^4}{(1-x^2)^2}\right).
\ee
As in the case of ABJM theory, the second term in \eqref{index-one-mono} does not contribute to the contour integral. Hence, as a result, we conclude that 
\be
I_{\ytableausetup{boxsize=0.3em}\ydiagram{1}\,\,\ydiagram{1}}^{(+)}\left(x, z_1, q \right)=\tilde{I}_k^\textrm{sp} \left(x, z_1, q \right).
\ee
It shows a perfect agreement of the gravity and field theory index in the sector with one monopole operator.
\section{GJV theory}\label{GJV}
In this section,
we turn to another class of AdS$_4$ solution with SU(3) symmetry and its field theory dual proposed by Guarino, Jafferis and Varela \cite{Guarino:2015jca}. The gravity solution, on the one hand, is the $\mathcal{N}=2$ SU(3) $\times$ U(1) invariant fixed point of $D=4$, $\mathcal{N}=8$ dyonic ISO(7) gauged supergravity, which can be uplifted on $S^6$ to  $\mathcal{N}=2$ AdS$_4$ solutions in massive type IIA supergravity \cite{Guarino:2015qaa, Guarino:2015vca}. Its field theory dual, on the other hand, is given by $D=3$ Chern-Simons-matter theory consisting of one U(N) gauge multiplet with a CS level $k$ and three adjoint chiral multiplets enjoying SU(3) flavor symmetry. 
This superconformal field theory can be obtained as an IR fixed point of RG-flow, which is triggered by the addition of Chern-Simons terms to UV $\mathcal{N}=8$ SYM on D2-branes. This RG flow is studied holographically in \cite{Guarino:2016ynd}.

For the GJV theory, the gravity free energy and the $S^3$-free energy at large $N$ are shown to precisely agree \cite{Guarino:2015jca}. The entropy of AdS$_4$ black hole in massive IIA supergravity was successfully reproduced from the computation of the topologically twisted index \cite{Hosseini:2017fjo, Benini:2017oxt}. Recently $\tau_{RR}$ was calculated in four-dimensional gauged supergravity \cite{Amariti:2021cpk}.
In this section, we compute the gravity index and the superconformal index at large $N$, and show that they also perfectly agree. 
\subsection{Gravity index}
For the AdS$_4$ solution in massive IIA supergravity, the Kaluza-Klein graviton spectrum  was studied in \cite{Pang:2017omp}.
Recently, the complete KK spectrum was obtained using the ExFT techniques \cite{Varela:2020wty}. In this section, we compute the graviton index. 
Since the analysis is parallel to the mABJM case presented in the previous section, we briefly
sketch the calculation and present the result. The SU(3) Dynkin labels, R-charges and the energy of the short multiplets are summarized in table \ref{table:GJV}. 
See table 2 in \cite{Varela:2020wty}.
\begin{table}[h!]
 \begin{center}
\begin{tabular}{l|ll||l}
 & \hfil $[p,q]_r$ & \hfil $E_0$ &\hfil index\\
\hline
 SGRAV & $[n,0]_{-2n/3} \oplus [0,n]_{2n/3}$ & $E_0=\frac{2n}{3}+2$ &
 $-x^{\frac{2(6+n)}{3}}$\\
 SGINO & $[n,1]_{-(2n+1)/3} \oplus [1,n]_{(2n+1)/3}$ & $E_0=\frac{11}{6}+\frac{2n}{3}$ &
 $\phantom{+}x^{\frac{2(5+n)}{3}}$\\
 SVEC & $[n+1,1]_{-2n/3} \oplus [1,n+1]_{2n/3}$ & $E_0=\frac{2n}{3}+1$ &
 $-x^{\frac{2(3+n)}{3}}$\\ 
 HYP & $[n+2,0]_{-2(n+2)/3} \oplus [0,n+2]_{2(n+2)/3}$ & $E_0=\frac{2n}{3}+\frac{4}{3}$ &
 $\phantom{+}x^{\frac{2(2+n)}{3}}$\\
\end{tabular}
\caption{The spectrum of the short multiplets at the Kaluza-Klein level $n$ and the contributions to the superconformal index for ${\mathcal{N}}=2$ SU(3) $\times$ U(1) invariant AdS$_4$ solution in massive IIA supergravity.}
\label{table:GJV}
\end{center} 
\end{table}

Here, we compute the contributions to the index of each supermultiplet and write them down in the last column of the table above. 
Multiplying the SU(3) character and summing over all the KK levels, we obtain the single-graviton index as
\begin{align}
&I_{\textrm{GJV}}^{\textrm{sp}}\left(x,y_1,y_2 \right)\nn\\
&=\dfrac{1}{1-x^2}\sum_{n=0}^{\infty}
\left(
- x^{\frac{2(6+n)}{3}}\, \chi_{SU(3)}^{[0,n]}
+ x^{\frac{2(5+n)}{3}} \, \chi_{SU(3)}^{[1,n]}
- x^{\frac{2(3+n)}{3}} \, \chi_{SU(3)}^{[1,n+1]}
+ x^{\frac{2(2+n)}{3}} \, \chi_{SU(3)}^{[0,n+2]}
\right)\nn\\
&\phantom{=}\,\,
+\dfrac{1}{1-x^2} \left(- x^{\frac{4}{3}} \, \chi_{SU(3)}^{[1,0]}
+ x^{\frac{2}{3}} \, \chi_{SU(3)}^{[0,1]} \right),\label{GJV-grav-sp}\\
&=-\dfrac{1}{1-x^2}-\dfrac{x^{\frac{2}{3}}}{x^{\frac{2}{3}}-y_1}
+\dfrac{x^{\frac{2}{3}}y_1}{y_2-x^{\frac{2}{3}}y_1}
+\dfrac{1}{1-x^{\frac{2}{3}}y_2}.
\end{align}
We have added the singletons contributions in the second line of equation \eqref{GJV-grav-sp}. It corresponds to the decoupled U(1) sector of U(N) GJV theory.
Evaluating the plethystic exponential \eqref{mp-sp}, we obtain the index of the multi-graviton as
\be\label{GJV-grav-mp}
I_{\textrm{mp}}
=\prod_{n=1}^{\infty}\dfrac{(1-x^{2n})}{
\left(1-{y_1}^{-n} x^{\frac{2}{3}n}\right)
\left(1-{y_1}^{n} {y_2}^{-n} x^{\frac{2}{3}n}\right)
\left(1- {y_2}^{n} x^{\frac{2}{3}n}\right)}.
\ee
\subsection{Superconformal index}
Now we move on to the calculation of the superconformal index of GJV theory.\footnote{The superconformal index of GJV theory with complex fugacities, which scales as $N^{\frac{5}{3}}$, was discussed in \cite{Bobev:2019zmz,Choi:2019dfu}.} The theory has three chiral multiplets $X_1, X_2$ and $X_3$ with the conformal dimensions $\frac{2}{3}$.
\begin{center}
\begin{tabular}{c|c c c}
 &  & GJV & \\
\hline
fields &  $t_1$ & $t_2$ & $\Delta$\\
\hline
$X_1$  & $0$             &   $\frac{1}{2}$   & $\frac{2}{3}$ \\ 
$X_2$  & $\frac{1}{2}$   &   $-\frac{1}{2}$   & $\frac{2}{3}$ \\ 
$X_3$  & $-\frac{1}{2}$  &   $0$   & $\frac{2}{3}$ \\ 
\end{tabular}
\end{center}
Here $t_1$ and $t_2$ are the charges of the Cartan generators of SU(3).
We easily read off the matrix $M$ \eqref{fprime} in
\begin{align}
&f^{'}_{\textrm{vector}} = -\lambda_{1,+1}\lambda_{1,-1},\nn\\
&f^{'}_{\textrm{chiral}} =\dfrac{\lambda_{1,+1}\lambda_{1,-1}}{1-x^2}\sum_{X_a}\left(z_i^{F_i} x^\Delta- z_i^{-F_i} x^{2-\Delta}\right),
\end{align}
and obtain the index as
\be
I^{(0)}=\prod_{n=1}^{\infty}\dfrac{1}{\textrm{det}M}
=\prod_{n=1}^{\infty}\dfrac{(1-x^{2n})}{
\left(1-{z_1}^{-\frac{n}{2}} x^{\frac{2n}{3}}\right)
\left(1-{z_1}^{\frac{n}{2}}{z_2}^{-\frac{n}{2}} x^{\frac{2n}{3}}\right)
\left(1- {z_2}^{\frac{n}{2}} x^{\frac{2n}{3}}\right)}.
\ee
Upon replacing
\be
z_1 \rightarrow y_1^2, \quad z_2 \rightarrow y_2^2,
\ee
one can easily see exact agreement with the gravity calculation in \eqref{GJV-grav-mp}.
\section{Conclusions}\label{discussion}
In this short note, we have discussed the large $N$ superconformal index for two non-trivial $\mathcal{N}=2$ Chern-Simons theories known as mABJM and GJV theories. They appear as IR-fixed points of the RG-flows driven by a superpotential mass deformation of UV ABJM theory and a Chern-Simons deformation of UV SYM on D2-branes, respectively. We have adopted the prescription provided in \cite{Kim:2009wb, Imamura:2011su} and computed the field theory index. We have also computed the graviton index on ${\mathcal{N}}=2$ SU(3) $\times$ U(1)-invariant AdS$_4$ solutions in D=11 and massive IIA supergravity, where we have extensively used the results of recent studies on the Kaluza-Klein spectrum on these backgrounds.  

Given the full KK spectrum, the computation of the graviton index is very straightforward and yields the expression in terms of SU(3) characters for both cases. However, the field theory computation for mABJM theory is rather subtle because the baryon-like U(1)$_b$ symmetry, under which the monopole operator is charged, is not a symmetry of mABJM theory. We can explicitly keep track of it at the expense of introducing one more fugacity related to U(1)$_b$ symmetry and follow the method of \cite{Kim:2009wb}. Then, we compute the index in the neutral sector and in the sector with one monopole operator. Something similar happens on the gravity side. To compare with the field theory index, one has to expand the graviton index \eqref{mABJM-gravity} in the fugacity associated with the monopole operator. However, we could not identify it in the graviton index. Hence, we have devised the so-called refined graviton index, by using the fact that ${\mathcal{N}}=2$ SU(3) $\times$ U(1)-invariant AdS$_4$ solutions in $D=11$ supergravity
is related to AdS$_4 \times$ S$^7$ solution by an RG-flow, and managed to identify it. As a result, we have successfully reproduced the field theory results.

For the GJV theory, there is no such issue. The gravity index is in perfect agreement with the field theory index. As a generalization of GJV theory, there is a large class of dual pairs of AdS$_4$ solutions in massive IIA supergravity and $D=3$ Chern-Simons theories with non-zero CS levels, which are inherited from the parent $D=4$, $\mathcal{N}=1$ superconformal field theory \cite{Fluder:2015eoa}. We expect that the computation of the field theory index is
a relatively easy task.
It would be very interesting to study the Kaluza-Klein spectrum of these background and compute the graviton index to give a concrete test of AdS$_4$/CFT$_3$ correspondence. 
\acknowledgments
We would like to thank Thomas Basile for helpful discussions. This work was supported by the National Research Foundation of Korea (NRF) grant 
2019R1A2C2004880(HK, NK) and 2020R1A2C1008497(HK).
\appendix
\section{Summary of the superconformal index of ABJM theory}
In this appendix, we summarize the gravity and the field theory index of ABJM theory discussed in \cite{Bhattacharya:2008zy, Bhattacharya:2008bja, Kim:2009wb}. 
\subsection{Gravity index}
The single-graviton index in $AdS_4 \times S^7$ \cite{Bhattacharya:2008zy} is given by
\begin{align}\label{ABJM-index}
I_{\textrm{ABJM}}^{\textrm{sp}}\left(x, y_1, y_2, y_3\right)
&=\textrm{Tr} [(-1)^F x^{\epsilon_0+j} y_1^{h_2} y_2^{h_3} y_3^{h_4}],\nn\\
&= \dfrac{1}{1-x^2}\left(\sum_{n=1}^{\infty} \left(x^{\frac{n}{2}}\,\chi_{SO(6)}^{(\frac{n}{2},\frac{n}{2},-\frac{n}{2})}-x^{\frac{n}{2}+1}\,\chi_{SO(6)}^{(\frac{n}{2},\frac{n}{2},-\frac{n-2}{2})} \right)\right.\\
&\phantom{= \dfrac{1}{1-x^2}\left( \right.}\left.+\sum_{n=2}^{\infty} \left(x^{\frac{n}{2}+2}\,\chi_{SO(6)}^{(\frac{n}{2},\frac{n-2}{2},-\frac{n-2}{2})}-x^{\frac{n}{2}+3}\,\chi_{SO(6)}^{(\frac{n-2}{2},\frac{n-2}{2},-\frac{n-2}{2})} \right)\right),\nn\\
&= \sum_{n=2}^{\infty} I_{R_n}+I_{R_1}
= \dfrac{\textrm{numerator}}{\textrm{denominator}},
\end{align}
where 
\begin{align}
\textrm{numerator}&=\sqrt{y_1 y_2 y_3}( y_1+y_2+y_3+y_1 y_2 y_3)x^{\frac{7}{2}}
-\sqrt{y_1 y_2 y_3}(1+ y_1 y_2+ y_2 y_3+y_3 y_1)x^{\frac{1}{2}}\nn\\
&\phantom{=}-( y_1 y_2+ y_2 y_3+y_3 y_1+y_1 y_2 y_3(y_1+y_2+y_3))(x^3-x),\\
\textrm{denominator}&= (1-x^2)(\sqrt{y_1}-\sqrt{x y_2 y_3})(\sqrt{y_2}-\sqrt{x y_3 y_1})(\sqrt{y_3}-\sqrt{x y_1 y_2})(\sqrt{x}-\sqrt{y_1 y_2 y_3}).\nn
\end{align}
The index in the large $k$ limit can be computed by
\begin{align}
I_{\textrm{ABJM}}^{\textrm{sp}}\left(x, y_1, y_2\right)
&=\int_c \dfrac{1}{2\pi i} \dfrac{d \sqrt{y_3}}{\sqrt{y_3}} I_{\textrm{ABJM}}^{\textrm{sp}}\left(x, y_1, y_2, y_3\right),\nn\\
&= \dfrac{x/y_1}{1 -x/y_1}+\dfrac{1}{1-x y_1}+\dfrac{x/y_2}{1 -x/y_2}+\dfrac{1}{1-x y_2}-\dfrac{2}{1-x^2},
\end{align}
where the contour includes the poles at $y_3=0, y_3=x y_1 y_2$ and $y_3=\frac{x}{y_1 y_2}$ \cite{Bhattacharya:2008bja}.
The index of single-graviton in $AdS_4 \times S^7/\mathbb{Z}_k$ for arbitrary $k$ was calculated in \cite{Kim:2009wb}.

The index of the multi-graviton is given by 
\be\label{mp-sp}
I^{\textrm{mp}}(x,y_1,y_2,y_3)= \textrm{exp} \left[\sum_{n=1}^\infty \dfrac{1}{n}
 I^{\textrm{sp}}(x^n,y_1^n,y_2^n,y_3^n)\right].
\ee
The multi-graviton index in the large $k$ limit becomes
\be\label{mp-abjm-0}
I_{\textrm{ABJM}}^{\textrm{mp}}\left(x, y_1, y_2\right)
=\prod_{n=1}^{\infty}\dfrac{(1-x^{2n})^2}{
\left(1-y_1^n x^n\right)
\left(1-y_1^{-n} x^n\right)
\left(1-y_2^n x^n\right)
\left(1-y_2^{-n} x^n\right)}.
\ee
The multi-graviton index with unit KK-momentum can be obtained by reading off the coefficient of  $y_3^{\frac{k}{2}}$ in $I_{\textrm{ABJM}}^{\textrm{mp}}\left(x, y_1, y_2, y_3\right)$ and given by  
$I_{k,\textrm{ABJM}}^{\textrm{sp}}\left(x, y_1, y_2\right)$, which can be calculated from the single-particle index as
\be\label{ABJM-gravity}
I_{k,\textrm{ABJM}}^{\textrm{sp}}\left(x, y_1, y_2\right)
= \oint
\dfrac{d \sqrt{y_3}}{(2\pi i)\sqrt{y_3}} y_3^{-\frac{k}{2}}
I_{\textrm{ABJM}}^{\textrm{sp}}\left(x, y_1, y_2, y_3\right).
\ee

One can rewrite the single-graviton index in terms of $SU(4)$ characters. Using the formula we have recorded in the appendix \ref{cha}, we obtain
\begin{align}\label{ABJM-gravity-su4}
(1-x^2)I_{\textrm{ABJM}}^{\textrm{sp}}&= \sum_{n=0}^{\infty} 
\left(- x^{\frac{n}{2}+4} \chi_{SU(4)}^{[n,0,0]}+x^{\frac{n}{2}+3}\chi_{SU(4)}^{[n,1,0]} 
-x^{\frac{n}{2}+2}\chi_{SU(4)}^{[n+1,0,1]}+ x^{\frac{n}{2}+1} \chi_{SU(4)}^{[n+2,0,0]}\right)\nn\\
&+ \left( - x^{\frac{3}{2}} \chi_{SU(4)}^{[0,0,1]}+ x^{\frac{1}{2}} \chi_{SU(4)}^{[1,0,0]}\right). 
\end{align}
The four terms in the first line are the contributions from the short graviton, short gravitino, short vector and hyper multiplet, respectively. The second line contributions come from the singletons. Evaluating the infinite sums, we obtain 
\begin{align}\label{ABJM-gravity-su4-sum}
I_{\textrm{ABJM}}^{\textrm{sp}}=
\dfrac{(1-x_1^{-1}x^{\frac{3}{2}})(1-x_2^{-1}x^{\frac{3}{2}})(1-x_3^{-1}x^{\frac{3}{2}})
(1- x_1 x_2 x_3 x^{\frac{3}{2}})}
{(1-x_1 x^{\frac{1}{2}})(1-x_2 x^{\frac{1}{2}})(1-x_3 x^{\frac{1}{2}})
(1-(x_1 x_2 x_3)^{-1}x^{\frac{1}{2}})(1-x^2)^2}
-\dfrac{1-x^2+x^4}{(1-x^2)^2}.
\end{align}
This result also can be reproduced by the method studied in \cite{Eager:2013mua}.\footnote{See equation (3.24) of \cite{Eager:2013mua}.}
\subsection{Superconformal index}
One can easily compute the neutral part of the index $I^{(0)}=\prod_{n=1}^{\infty}\left(1/\textrm{det}M\right)$ and show that it exactly agrees with the gravity calculation \eqref{mp-abjm-0}
\cite{Bhattacharya:2008bja}.
The index with unit KK momentum was calculated in \cite{Kim:2009wb} and given by
\be\label{ABJM-mono}
I_{\ytableausetup{boxsize=0.3em}\ydiagram{1}\,\,\ydiagram{1}}^{(+)}
= \oint
\dfrac{dz}{(2\pi i)z} z^{-k}
\left[\dfrac{
\left(1-x \sqrt{x y_1 y_2} z\right)
\left(1-x \sqrt{\frac{x }{y_1 y_2}} z\right)
\left(1-x \sqrt{\frac{x y_1}{ y_2}} z^{-1}\right)
\left(1-x \sqrt{\frac{x y_2 }{y_1}} z^{-1}\right)}{
\left(1-\sqrt{\frac{x y_1}{ y_2}} z \right)
\left(1-\sqrt{\frac{x y_2}{y_1 } z} \right)
\left(1-\sqrt{x y_1 y_2} z^{-1} \right)
\left(1-\sqrt{\frac{x }{y_1 y_2}} z^{-1} \right)
\left(1-x^2\right)^2}\right].
\ee
Identifying $z \equiv\sqrt{y_3}$, one can obtain
\be\label{ABJM-field}
I_{\ytableausetup{boxsize=0.3em}\ydiagram{1}\,\,\ydiagram{1}}^{(+)}
= \oint
\dfrac{d \sqrt{y_3}}{(2\pi i)\sqrt{y_3}} y_3^{-\frac{k}{2}}
\left[
I_{\textrm{ABJM}}^{\textrm{sp}}\left(x, y_1, y_2, y_3\right)+ 
\dfrac{1-x^2+x^4}{(1-x^2)^2}
\right].
\ee
Comparing the gravity result \eqref{ABJM-gravity} and the field theory result \eqref{ABJM-field}, one can conclude that
\be
I_{k,\textrm{ABJM}}^{\textrm{sp}}\left(x, y_1, y_2\right)=
I_{\ytableausetup{boxsize=0.3em}\ydiagram{1}\,\,\ydiagram{1}}^{(+)}\left(x, y_1, y_2\right).
\ee
It shows that the gravity index and the field theory index in the sector with unit KK momentum exactly agree.
\section{Character formulas}\label{cha}
In the main text, we have been dealing with SU(3) character to compute the single-graviton index. The SU(3) character with Dynkin labels $[M_1, M_2]$ is given by
\begin{align}\label{chi-su(3)}
\chi_{SU(3)}^{M_1,M_2} (x_1,x_2,x_3)
=\dfrac{\textrm{det}x_i^{n_j+3-j}}{\textrm{det}(x_i^{3-j})}=
\begin{vmatrix}
 x_1^{M_1+M_2+2} &x_1^{M_2+1}&1\\
 x_2^{M_1+M_2+2} &x_2^{M_2+1}&1\\
 x_3^{M_2+M_2+2} &x_3^{M_2+1}&1
\end{vmatrix} 
\Bigg /
\begin{vmatrix}
 x_1^2 & x_1 & 1\\
 x_2^2 & x_2 & 1\\
 x_3^2 & x_3 & 1
\end{vmatrix}
,
\end{align}
where 
\be
n_j= \sum_{i=j}^2 M_i \quad \textrm{for}\, j=1,2, \quad n_3=0,
\ee
and $x_i$ satisfy $x_1 x_2 x_3 =1$. We have used
\be
x_1=y_1, \quad x_2=\dfrac{1}{y_2}, \quad x_3= \dfrac{y_2}{y_1}.
\ee

For the ABJM theory, we need the SU(4) or SO(6) character to compute the gravity index. The expression to calculate the SU(4) character is straightforward by generalizing the above formula. The $SO(6)$ character with the highest weights $\lambda_1 \geq \lambda_2 \geq|\lambda_3| \geq 0$ is given by
\begin{align}
\chi_{SO(6)}^{\lambda_1,\lambda_2,\lambda_3} (y_1,y_2,y_3)=
\dfrac{\textrm{det}[\textrm{sinh}(y_i(\lambda_i+3-j))]+\textrm{det}[\textrm{cosh}(y_i(\lambda_i+3-j))]}{\textrm{det}[\textrm{cosh}(y_i(3-j))]}.
\end{align}
Dynkin labels $[M_1,M_2,M_3]$ is related to highest weight labels $\lambda_1, \lambda_2, \lambda_3$ by $M_1= \lambda_2-\lambda_3,~ M_2=\lambda_1-\lambda_2$ and $M_3=\lambda_2+\lambda_3$. 
Using the fugacity map
\be
x_1^2= \dfrac{y_2 y_3}{y_1}, \quad x_2^2= \dfrac{y_1 y_3}{y_2}, \quad x_3^2= \dfrac{y_1 y_2}{y_3},  \quad x_4^2 =\dfrac{1}{y_1 y_2 y_3},
\ee
one can explicitly check that
\begin{align}
\chi_{SO(6)}^{(\frac{n}{2},\frac{n}{2},-\frac{n}{2})} =\chi_{SU(4)}^{[n,0,0]}, \quad
\chi_{SO(6)}^{(\frac{n}{2},\frac{n}{2},-\frac{n-2}{2})} =\chi_{SU(4)}^{[n-1,0,1]} \quad
\textrm{for}\quad n \geq 1, \\
\chi_{SO(6)}^{(\frac{n}{2},\frac{n-2}{2},-\frac{n-2}{2})} =\chi_{SU(4)}^{[n-2,1,0]}, \quad
\chi_{SO(6)}^{(\frac{n-2}{2},\frac{n-2}{2},-\frac{n-2}{2})} =\chi_{SU(4)}^{[n-2,0,0]} \quad
\textrm{for}\quad n \geq 2.
\end{align}
\section{A brief review on computing the $\mathcal{N}=2$ superconformal index}\label{review-SCI}
In this appendix, we summarize some formulas of \cite{Imamura:2011uj}, which are needed to compute the superconformal index at large $N$ in the main text. For details, we refer to \cite{Kim:2009wb, Imamura:2011su}.
In the large $N$ limit, the index is factorized into the following three parts
\be
I= I^{(0)}I^{(+)}I^{(-)},
\ee
where $I^{(0)}$ is independent of the magnetic charges and $I^{(+)} \left(I^{(-)}\right)$ depends only on the positive (negative) magnetic charges. Similarly, the letter index $f=f_{\textrm{vector}}+f_{\textrm{chiral}}$ is also decomposed into $f'+f^{(+)}+f^{(-)}$. 
Then, the neutral part of the index can be easily calculated as
\be\label{neutral-index}
I^{(0)}
=\prod_{n=1}^{\infty}
\dfrac{1}{\textrm{det}M({z_i}^n, x^n)},
\ee
where the matrix $M$ can be read off from $f'$
\begin{align}\label{fprime}
f'
&=-\sum_{A=1}^{n_G} \lambda_{A,+1} \lambda_{A,-1}
+\sum_{\Phi_{A,B}} \left[ \lambda_{A,+1}\lambda_{B,-1} \dfrac{z_i^{F_i}x^{\Delta(\Phi)}}{1-x^2}
-\lambda_{A,-1}\lambda_{B,+1} \dfrac{z_i^{-F_i}x^{2-\Delta(\Phi)}}{1-x^2}\right],\\
&=-\sum_{A,B} \lambda_{A,+1} M_{A, B} \lambda_{B,-1}.
\end{align}
The index $I^{(+)}$, which is the contribution from the positive magnetic charges, is given by
\be
I^{(+)}=\sum_{m^{(+)}}\int da^{(+)} e^{-S_{CS}^{(+)}} x^{\epsilon_0^{(+)}} z_i^{q_{0i}^{(+)}}
\textrm{exp}\left[ \sum_{n=1}^\infty \dfrac{1}{n} f^{(+)}\left(e^{i n a}, x^n, z_i^n \right) \right].
\ee
The contributions from Chern-Simons term, the zero-point energy and the zero-point flavor charges can be calculated through 
\begin{align}
S_{CS}^{(+)}&= i \sum_{A=1}^{n_G} \sum_{\rho \in \mathbf{N}_A}^{(+)} k_A\, \rho(a)\, \rho(m),\\
\epsilon_0^{(+)}&= \dfrac{1}{2} \sum_{\Phi_{AB}} \sum_{\rho \in \mathbf{N}_A}^{(+)}\sum_{\rho' \in \mathbf{N}_B}^{(+)} \left(\left|\rho(m)-\rho'(m)\right|-\left|\rho(m)\right|-\left|\rho'(m)\right| \right) \left(1-\Delta(\Phi) \right)\nn\\
&\phantom{=}- \dfrac{1}{2} \sum_{A=1}^{n_G} \sum_{\rho \in \mathbf{N}_A}^{(+)}\sum_{\rho' \in \mathbf{N}_A}^{(+)} \left(\left|\rho(m)-\rho'(m)\right|-\left|\rho(m)\right|-\left|\rho'(m)\right| \right),\\
q_{0i}^{(+)}&= -\dfrac{1}{2} \sum_{\Phi_{AB}} \sum_{\rho \in \mathbf{N}_A}^{(+)}\sum_{\rho' \in \mathbf{N}_B}^{(+)} \left(\left|\rho(m)-\rho'(m)\right|-\left|\rho(m)\right|-\left|\rho'(m)\right| \right) F_i (\Phi).
\end{align}
The letter indices for vector and chiral multiplets, which are associated to the positive magnetic charges, are given by
\begin{align}
f_{\textrm{vector}}^{(+)}\label{f-vec}
&=  \sum_{A=1}^{n_G} \sum_{\rho \in \mathbf{N}_A}^{(+)}\sum_{\rho' \in \mathbf{N}_A}^{(+)}
\left[-\left(1-\delta_{\rho,\rho'}\right) x^{|\rho(m)-\rho'(m)|}+x^{|\rho(m)|+|\rho'(m)|}\right]e^{i \left(\rho(a)-\rho'(a) \right)},\\
f_{\textrm{chiral}}^{(+)}\label{f-chi}
&=  \sum_{\Phi_{AB}}\sum_{\rho \in \mathbf{N}_A}^{(+)}\sum_{\rho' \in \mathbf{N}_B}^{(+)} 
\left( x^{|\rho(m)-\rho'(m)|}-x^{|\rho(m)|+|\rho'(m)|} \right)\nn\\
&\phantom{=} \times \dfrac{1}{1-x^2} \left(e^{i\left(\rho(a)-\rho'(a) \right)} z_i^{F_i(\Phi)} x^{\Delta(\Phi)}-e^{-i\left(\rho(a)-\rho'(a) \right)} z_i^{-F_i(\Phi)} x^{2-\Delta(\Phi)}\right).
\end{align} 






\bibliography{KK}{}
\end{document}